# Dual-Comb Spectroscopic Polarimetry for Full Jones Matrix Retrieval Enabled by Polarization-Controlled Pulse Sequences

**Hidenori Koresawa,[a] Hiroki Kitahama,[b] Shogo Tanimura,[b] Eiji Hase,[c,d] Yu Tokizane,[c,d] Akifumi Asahara,[e] Takeo Minamikawa,[c,f] Kaoru Minoshima,[e] and Takeshi Yasui[c,d,*]**

[a]Tokushima University, Graduate School of Advanced Technology and Science, 2-1 Minami-Josanjima, Tokushima, Japan

[b]Tokushima University, Graduate School of Sciences and Technology for Innovation, 2-1 Minami-Josanjima, Tokushima, Japan

[c]Tokushima University, Institute of Post-LED Photonics (pLED), 2-1 Minami-Josanjima, Tokushima, Japan

[d]Tokushima University, Institute of Photonics and Human Health Frontier (IPHF), 2-1, Minami-Josanjima, Tokushima, Japan

[e]The University of Electro-Communications, Graduate School of Informatics and Engineering, 1-5-1 Chofugaoka, Chofu, Japan

[f]The University of Osaka, Graduate School of Engineering Science, 1-3 Machikaneyama, Toyonaka, Japan

**Abstract**. Spectroscopic polarimetry (SP) is a valuable technique for evaluating thin films, optical materials, and biological samples by revealing both polarimetric and spectroscopic properties. However, its performance is limited by mechanical instability and slow data acquisition due to polarization modulation. To address this, we combine dual-comb spectroscopic polarimetry (DCSP) with polarization-controlled pulse sequences (PCPS) having distinct polarizations and time delays. This approach enables detailed characterization of a sample's polarization response using the Jones matrix, providing both real and imaginary components as a function of wavelength by simultaneously measuring optical amplitude and phase spectra in two orthogonal polarizations. This method, termed Jones matrix DCSP (JM-DCSP), allows for accurate analysis of optical elements with known polarization properties, with experimental results in good agreement with theoretical predictions. By eliminating the need for mechanical modulation and enabling multiplexed polarization probing, JM-DCSP enhances the precision and utility of SP across a broad spectral range and may be applied to various optical characterization scenarios.

***Takeshi Yasui, E-mail: yasui.takeshi@tokushima-u.ac.jp**

## 1 Introduction

Spectroscopic polarimetry (SP), also known as spectroscopic ellipsometry,[1] is a well-established technique that combines spectroscopy and polarimetry to investigate the wavelength-dependent polarization properties of light. This enables detailed analysis of light-matter interactions, including phenomena such as optical activity, birefringence, and circular dichroism. Consequently, SP plays a crucial role in evaluating thin films,[2] characterizing optical materials,[3] and analyzing biological specimens.[4]

Usual SPs measure broad spectra of a phase difference $\Delta$ and an amplitude ratio $\psi$ between two orthogonally linear-polarized lights (x-polarization component and y-polarization component in transmission configuration or p-polarization component and s-polarization component in oblique-incidence reflection configuration) when the broadband light with known polarization is incident onto the sample. To obtain these spectra, a spectrum of optical intensity is measured while modulating the polarization of the incident or the output light, and the resulting spectrum is used for determining the spectra of $\Delta$ and $\psi$. Depending on methods of polarization modulation, SP can be classified into three types: rotating-analyzer SP,[5] rotating-compensator SP,[6] and phase modulation SP.[7] The rotating-analyzer SP and the rotating-compensator SP acquire the spectra of $\Delta$ and $\psi$ by a combination of a multi-channel spectrometer with a mechanically rotating polarization optics. Although these SPs have own advantages such as easy implementation and moderate precision, such mechanical polarization modulation limits the mechanical stability and the data acquisition time (typically, several tens of milliseconds) of the SP system. On the other hand, the phase modulation SP benefits from non-mechanical, rapid polarization modulation with a photoelastic modulator (PEM), and hence can reduce the data acquisition time (typically, several tens of microseconds). However, since phase modulation with PEM shows a large dependent on the wavelength and the temperature, the precise compensation of phase modulation is required for SP. Furthermore, the fast modulation speed with PEM (tens of kilohertz) is not so good compatible with the use of a multi-channel spectrometer equipped with a camera (frame rate = 10 ~ 1000 fps). These features make it difficult to apply PEM for broadband SP.

Recently, optical frequency comb (OFC) has attracted attention as a light source for polarization-modulation-free SP.[8-11] OFC[12-14] can be perceived as an ensemble of tens of thousands of individually phase-locked single-wavelength lasers with equal frequency intervals (= $f_{rep}$), combining a broadband spectral characteristic with narrow linewidth mode properties. Also, dual-comb spectroscopy (DCS)[15-18] enables rapid, precise acquisition of mode-resolved OFC spectra of optical amplitude and phase by interfering two OFCs with slightly different $f_{rep}$ to generate a secondary frequency comb of optical beat signals in RF region (namely, RF comb). In dual-comb spectroscopic polarimetry (DCSP) based on a combination of SP with DCS,[8-11] optical spectra of amplitude ratio $\psi$ and phase difference $\Delta$ can be obtained without polarization modulation by acquiring OFC spectra of optical amplitude and phase in two orthogonal polarization components of the output light to the incident light with known polarization. The resulting polarization-modulation-free characteristic does not suffer from the instability of polarization modulation and improve the performance of SP. Effectiveness of DCSP was demonstrated by applying it for analysis of a thin film[8,10] and a time-resolved polarization measurement of a dynamic sample.[9,10]

In the previous studies of DCSP, the polarization analysis of the output light from the sample in response to incident light with a single polarization was performed. However, if the incident light is multiplexed into multiple polarizations instead of using just a single polarization, it enables a more detailed analysis of the sample's polarization response. In other words, the use of multiple polarizations in the incident light can significantly enhance the performance of DCSP. For example, if two different polarizations are used for the incident light of DCSP, Jones matrix of a sample can be acquired as a function of wavelength. The Jones matrix is a 2×2 matrix that relates the input polarization state of light to the output polarization state after passing through a sample; it is often used to describe the polarization state of light as it propagates through the sample. The elements of the Jones matrix represent the complex amplitude transfer coefficients between two orthogonal polarization basis states. While we adopt the linear basis in this work for convenience, any orthonormal polarization basis (e.g., right/left circular[19)] or elliptical[20)]) can be used in principle, with the Jones matrix in different bases related by a unitary similarity transformation. To obtain the Jones matrix, polarization analysis of the output light for two different incident polarizations is required. Fortunately, due to the unique characteristics of OFC, the mode locking of multiple optical frequency modes in the frequency domain can be observed as ultra-short light pulses in the time domain; this suggests that it is possible to discretely multiplex different polarization states for multiple pulse sequences.

In this article, we developed Jones matrix DCSP (JM-DCSP). Polarization control pulse sequences (PCPS) with distinct polarizations and time delays are generated by separating a single optical pulse of OFC into different optical paths and then adding different time delay and the polarization control on each of them. The DCSP of PCPS enables Jones matrix SP without the need for polarization modulation, benefiting from the rapid, precise, data acquisition. We measured polarization elements with known polarization property to demonstrate the effectiveness of JM-DCSP.

## 2 Materials and Methods

### *2.1 Principle of Jones Matrix Reconstruction*

We describe the method for deriving the Jones matrix using JM-DCSP. Jones calculus is outlined in the Apendix section. In this article, we define x- and y-polarization as horizontal (or 0°) polarization and vertical (or +90°) polarization, respectively. A pair of OFC pulses is used for the generation of the first polarization-controlled pulse (PCP) and the second PCP, respectively. The first PCP maintains polarization that is not parallel to x- and y-polarization, whereas the second PCP has a different polarization and time delay from the first PCP. The polarization states of the first PCP and the second PCP before passing through the sample are represented as the Jones vectors, $\boldsymbol{E_1}$ and $\boldsymbol{E_2}$, given by

$$\boldsymbol{E_1} = \begin{bmatrix} E_{1x} \cdot e^{i\delta_{1x}} \\ E_{1y} \cdot e^{i\delta_{1y}} \end{bmatrix}, \tag{1}$$

$$\boldsymbol{E_2} = \begin{bmatrix} E_{2x} \cdot e^{i\delta_{2x}} \\ E_{2y} \cdot e^{i\delta_{2y}} \end{bmatrix}, \tag{2}$$

where $E_{1x}$ and $E_{1y}$ represent the amplitudes of the x- and y-polarized components, while $\delta_{1x}$ and $\delta_{1y}$ represent their respective phases in the first PCP. Similarly, $E_{2x}$ and $E_{2y}$ represent the amplitudes of the x- and y-polarized components, with $\delta_{2x}$ and $\delta_{2y}$ representing their phases in the second PCP. Conversely, the polarization states of the two PCPs after passing through the sample are represented as the Jones vectors, $\boldsymbol{E_1'}$ and $\boldsymbol{E_2'}$, by

$$\boldsymbol{E_1}' = \begin{bmatrix} E_{1x}' \cdot e^{i\delta_{1x}'} \\ E_{1y}' \cdot e^{i\delta_{1y}'} \end{bmatrix}, \tag{3}$$

$$\boldsymbol{E_2}' = \begin{bmatrix} E_{2x}' \cdot e^{i\delta_{2x}'} \\ E_{2y}' \cdot e^{i\delta_{2y}'} \end{bmatrix}, \tag{4}$$

where $E_{1x}'$ and $E_{1y}'$ represent the amplitudes, and $\delta_{1x}'$ and $\delta_{1y}'$ represent the phases of the x- and y-polarized components in the first PCP. Similarly, $E_{2x}'$ and $E_{2y}'$ denote the amplitudes, while $\delta_{2x}'$ and $\delta_{2y}'$ denote the phases of the x- and y-polarized components in the second PCP. From those four equations, Eqs. (5) and (6) are given by

$$\boldsymbol{E_1}' = \begin{bmatrix} J_{11} & J_{12} \\ J_{21} & J_{22} \end{bmatrix} \cdot \begin{bmatrix} E_{1x} \cdot e^{i\delta_{1x}} \\ E_{1y} \cdot e^{i\delta_{1y}} \end{bmatrix} = \begin{bmatrix} E_{1x}' \cdot e^{i\delta_{1x}'} \\ E_{1y}' \cdot e^{i\delta_{1y}'} \end{bmatrix}, \tag{5}$$

$$\boldsymbol{E}_2' = \begin{bmatrix} J_{11} & J_{12} \\ J_{21} & J_{22} \end{bmatrix} \cdot \begin{bmatrix} E_{2x} \cdot e^{i\delta_{2x}} \\ E_{2y} \cdot e^{i\delta_{2y}} \end{bmatrix} = \begin{bmatrix} E_{2x}' \cdot e^{i\delta_{2x}'} \\ E_{2y}' \cdot e^{i\delta_{2y}'} \end{bmatrix}, \tag{6}$$

where $J_{11}$, $J_{12}$, $J_{21}$, and $J_{22}$ are the complex elements of the Jones matrix. Ultimately, the elements of the Jones matrix can be derived by

$$\begin{bmatrix} J_{11} \\ J_{12} \end{bmatrix} = \begin{bmatrix} E_{1x} \cdot e^{i\delta_{1x}} & E_{1y} \cdot e^{i\delta_{1y}} \\ E_{2x} \cdot e^{i\delta_{2x}} & E_{2y} \cdot e^{i\delta_{2y}} \end{bmatrix}^{-1} \cdot \begin{bmatrix} E_{1x}' \cdot e^{i\delta_{1x}'} \\ E_{2x}' \cdot e^{i\delta_{2x}'} \end{bmatrix} \tag{7}$$

$$\begin{bmatrix} J_{21} \\ J_{22} \end{bmatrix} = \begin{bmatrix} E_{1x} \cdot e^{i\delta_{1x}} & E_{1y} \cdot e^{i\delta_{1y}} \\ E_{2x} \cdot e^{i\delta_{2x}} & E_{2y} \cdot e^{i\delta_{2y}} \end{bmatrix}^{-1} \cdot \begin{bmatrix} E_{1y}' \cdot e^{i\delta_{1y}'} \\ E_{2y}' \cdot e^{i\delta_{2y}'} \end{bmatrix} \tag{8}$$

### *2.2 Experimental setup*

Based on Eqs. (7) and (8), which provide the basis for calculating the Jones matrix, we constructed an experimental setup for JM-DCSP. Figure 1 shows a schematic drawing of experimental setup for JM-DCSP. A pair of mode-locked erbium-doped fiber OFCs (Neoark Co., Japan, OCLS-HSC-D100-TKSM, center wavelength = 1562 nm, spectral bandwidth = 50 nm) was used for a signal OFC ($f_{rep1}$ = 100 MHz, $f_{ceo1}$ = 10.5 MHz) and a local OFC ($f_{rep2}$ = 99.99968 MHz, $f_{ceo2}$ = 10.5 MHz) in JM-DCSP. $f_{rep1}$, $f_{ceo1}$, and $f_{ceo2}$ of these OFCs are phase-locked to a rubidium frequency standard (Rb-FS, Stanford Research Systems, Inc., FS725, frequency = 10 MHz, accuracy = 5×$10^{-11}$, instability = 2×$10^{-11}$ at 1 s). The local OFC was tightly locked to the signal OFC laser by finely adjusting its repetition rate ($f_{rep2}$) with a constant frequency offset $\Delta f_{rep}$ (= $f_{rep1}$ - $f_{rep2}$ = 320 Hz) using a narrow-linewidth CW laser (Redfern Integrated Optics, Inc., Santa Clara, CA, USA, PLANEX, center wavelength: 1,550 nm; FWHM: <2.0 kHz) for an intermediate laser.[21,22] This enables us to coherently accumulate interferograms obtained with JM-DCSP and hence to enhance SNR.[23,24]

An optical pulse of a signal OFC (OFC pulse) was split into a transmitted OFC pulse for PCPS (see red line, x-polarization) and a reflected OFC pulse for a reference pulse (see purple line, y-polarization) after adjusting their power split ratio of 1:2 by a combination of a quarter-wave plate (λ/4), a half-wave plate (λ/2), and a polarization beam splitter (PBS1). The reflected OFC pulse was used to determine the absolute phase of the first and second PCPs as well as give a trigger signal of JM-DCSP. Conversely, the transmitted OFC pulse was further split into the first PCP (see orange line) and the second PCP (see green line) by a beam splitter (BS1). The polarization of the first and the second PCPs was adjusted to be x-polarization and y-polarization while maintaining the similar optical power by a pair of λ/2 and a PBS2. Difference between their optical pathlengths causes a time delay of 0.63 ns between the first and second PCPs. Then, the first PCP and the second PCP were spatially overlapped again by PBS2, resulting in generation of PCPS. As the following λ/2 rotates the polarization of PCPS by +45°, the polarization before a sample was set to be +45° for the first PCP and +135° for the second PCP. The PCPS passed through the sample, and each pulse undergoes changes in polarization condition reflecting the sample's optical properties. The reference pulse passed through a polarizer (P, polarization angle = +45°) and then was combined with the PCPS by BS2, generating a train of three pulses, composed of the first PCP (+45° linear polarization), the second PCP (+135° linear polarization, time delay of 0.63 ns from the first PCP), and the reference pulse (+45° linear polarization, time delay of 0.15 ns from the second PCP). This pulse train triple, which is multiplexed in polarization and time, was fed into the optical setup for DCSP. The optical systems for the first PCP, the second PCP, and the reference pulse were enclosed in a plastic box to suppress the influence of environmental disturbances such as atmospheric turbulence.

The local OFC has +45° linear polarization after passing through λ/4, λ/2, and P; then, it was spatially overlapped with the combined PCPS and reference pulse by BS3 to generate the interferogram. An optical bandpass filter (BPF, center wavelength = 1560±2 nm, FWHM = 12±2.4 nm) was used to limit the spectrum to avoid detector saturation and enhance the measurement SNR. However, by relaxing the current spectral filtering conditions and utilizing photodetectors with higher dynamic range, the full anti-aliasing bandwidth of the system (≈ 15 THz for $\Delta f_{rep}$ = 320 Hz and $f_{rep1}$ = 100 MHz) could be exploited. In this case, the use of broadband mode-locked fiber combs with wider gain spectra, or the application of highly nonlinear fibers or optical waveguides for supercontinuum generation, would be highly effective. This suggests a potential for extending to broadband spectral measurement. The x- and y-polarization components of the interferogram were respectively detected by a combination of PBS3 and a pair of photodiodes (PDs), and then acquired by a digitizer (not shown).

### *2.3 Data analysis*

The acquired signal is an interferogram sequence generated by the first PCP, the second PCP, and the reference pulse. For each of these interferograms, only the center burst signal is temporally extracted, and zero padding is applied to the remaining data outside of the extracted portion. This process allows the separation of the interferograms generated by the first PCP, second PCP, and reference pulse, each within a time window of $1/f_{rep1}$. The Fourier transform of the zero-padded interferogram generated OFC spectra with amplitude and phase information for the x- and y-polarization components. We achieved a display spectral resolution of $f_{rep1}$, representing an interpolated resolution due to zero-padding, which does not enhance the actual physical spectral resolution. The phase difference spectra between the first PCP (or the second PCP) and the reference pulse were used to calculate the absolute phase spectra of the first PCP (or the second PCP). These spectra are then utilized in the Jones calculus to obtain the Jones matrix of the sample.

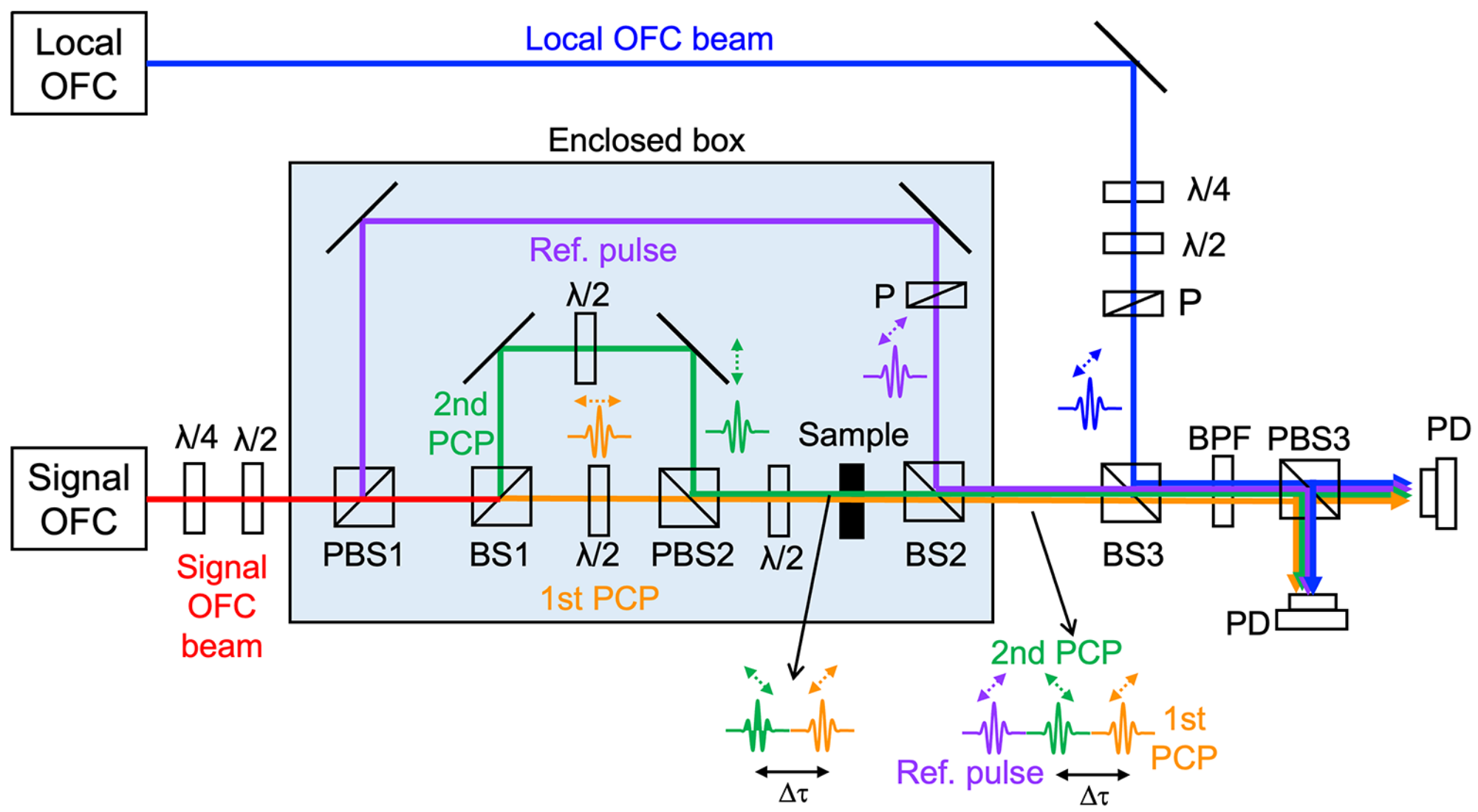

Fig. 1. Experimental setup of JM-DCSP. Rb-FS, rubidium frequency standard; λ/4, quarter-wave plate; λ/2, a half-wave plate; PBSs, polarization beam splitters; BSs, beam splitters; P, polarizer; BPF, optical bandpass filter; PDs, photodiodes.

## 3 Results

### *3.1 Basic Performance of JM-DCSP System*

First, the basic characteristics of the developed JM-DCSP system were evaluated. Figure 2 shows the temporal waveform of the interferogram obtained without a sample. A sequence of three interferograms can be observed as impulse signals within the effective time range of 0 ns to 1 ns. For achieving higher spectral resolution, it is ideal to appropriately disperse the sequence of three interferograms within a total time window of 10 ns. However, this requires a long optical path length difference between each pulse, which increases the influence of environmental disturbances and compromises the stability of the interferograms. Therefore, in this experiment, the time delay was set to a level where the influence of environmental disturbances would not be problematic. The three insets in Fig. 2 show the temporally magnified interferograms of the first PCP, the second PCP, and the reference pulse, respectively, where burst waveforms with an adequate signal-to-noise ratio (SNR) can be observed.

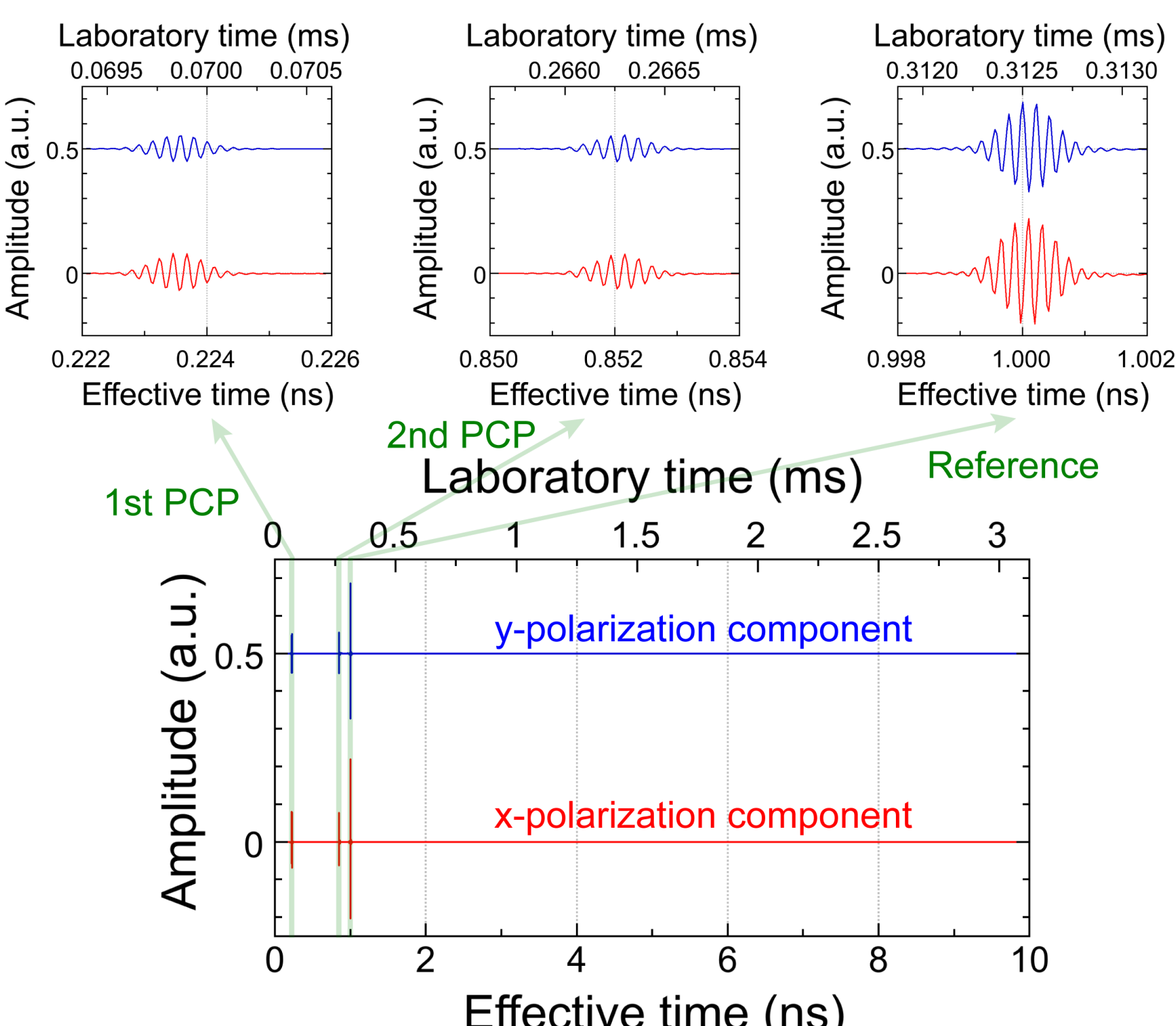

Fig. 2. Interferogram of no samples with respect to x-polarization and y-polarization components in one repetition period of the interferogram (= 10 ns). Three insets show the temporally magnified interferograms of the 1st PCP, 2nd PCP, and the reference pulse, respectively.

We obtained the amplitude spectra of the x- and y-polarization components of the first PCP, the second PCP, and the reference pulse by Fourier transforming, as shown in Figs. 3(a), 3(b), and 3(c), respectively. The spectral bandwidth was limited by the bandpass filter, and similar spectral shapes are obtained for both polarization components in them. The periodic spectral modulation superimposed on the amplitude spectrum is attributed to the spectral characteristics of the OFC sources and is caused by weak internal reflection echo signals generated by the component forming the fiber OFC cavity. The phase spectra of the x-polarization components of the first PCP, the second PCP, and the reference pulse are shown in Fig. 3(d). Difference of phase slope is corresponding to that of temporal position of each pulse in the interferogram signal (see Fig. 2). The phase difference spectra of the x- and y-polarization components between the first PCP and the reference pulse, as well as those between the second PCP and the reference pulse, were calculated, as shown in Figs. 3(e) and 3(f), respectively. These phase difference spectra show good overlap between x- and y-polarization components in them, again. Such good equivalence between the x- and y-polarization components enables accurate JM-DCSP.

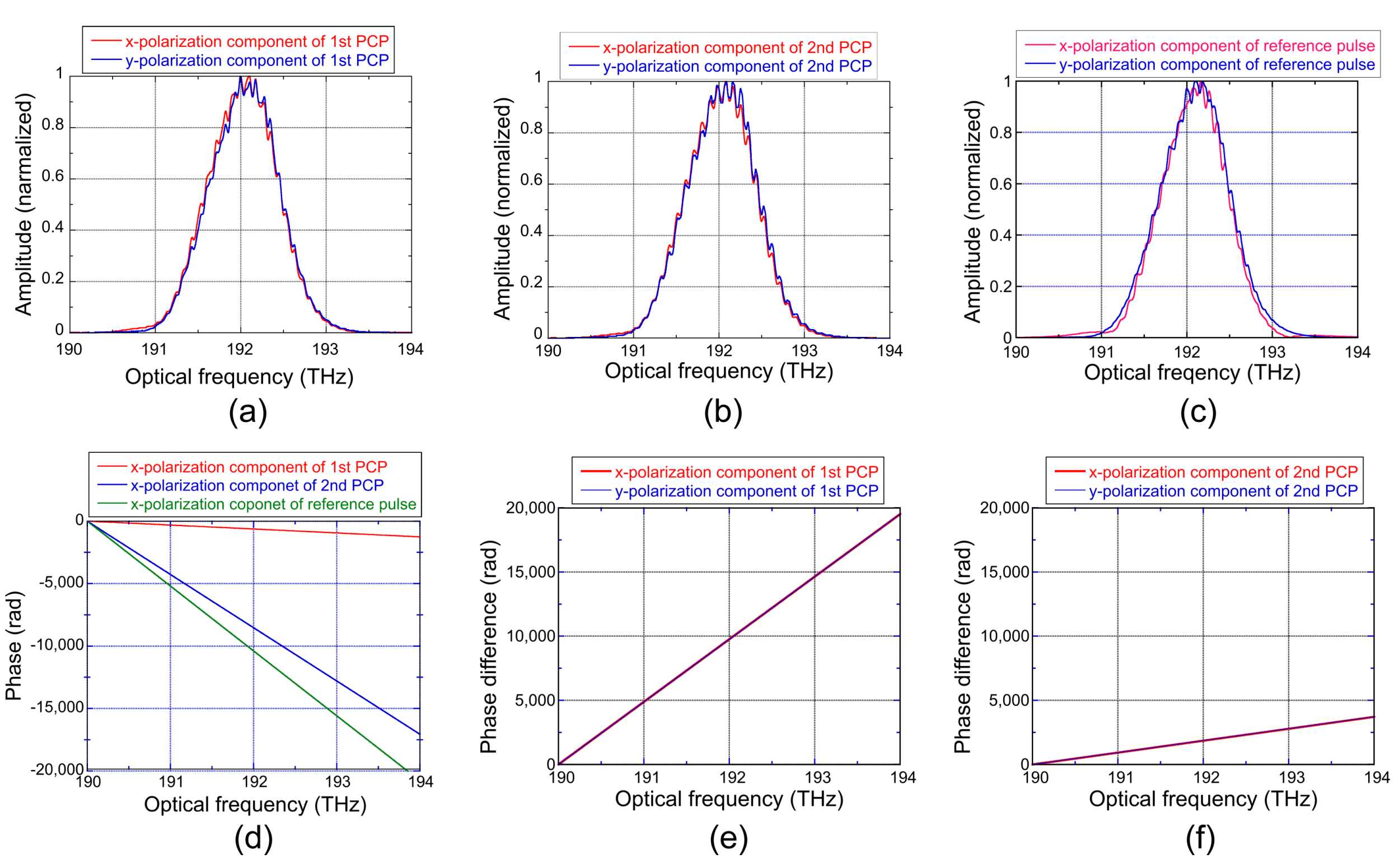

Fig. 3. Spectral characteristics of the first PCP, the second PCP, and the reference pulse. Amplitude spectra of the x- and y-polarization components of (a) the first PCP, (b) the second PCP, and (c) the reference pulse. (d) Comparison of phase spectra of the x-polarization component among the first PCP, the second PCP, and the reference pulse. Phase difference spectra of the x- and y-polarization components of (e) the first PCP and (f) the second PCP to the reference pulse.

We next evaluated the basic performance of PCPS. Red and blue traces of Fig. 4(a) show interferograms of no samples with respect to x- and y-polarization components. It is important to note that both x-polarized component and y-polarized component are moderately present in each interferogram because we utilize linear polarization directed at +45º or +135º for the PCPS, the reference pulse, and the local OFC pulse, again. The first PCP and the second PCP were observed with a time delay of 0.63 ns whereas the second PCP and the reference pulse were observed with a time delay of 0.15 ns. In order to verify the validity of their configured polarizations, we next placed a polarizer at the sample position. The red and blue traces in Fig. 4(b) represent the x- and y-polarization components of the interferogram when the placed polarizer was oriented with its transmission axis set at +45°. The first PCP and the reference pulse appeared whereas the second PCP disappeared as expected. Conversely, when the placed polarizer was oriented with its transmission axis set at +135°, the second PCP and the reference pulse appeared whereas the first PCP disappeared, as shown in Fig. 4(c). In this way, we verified the validity of the polarization for the first PCP, the second PCP, and the reference pulse.

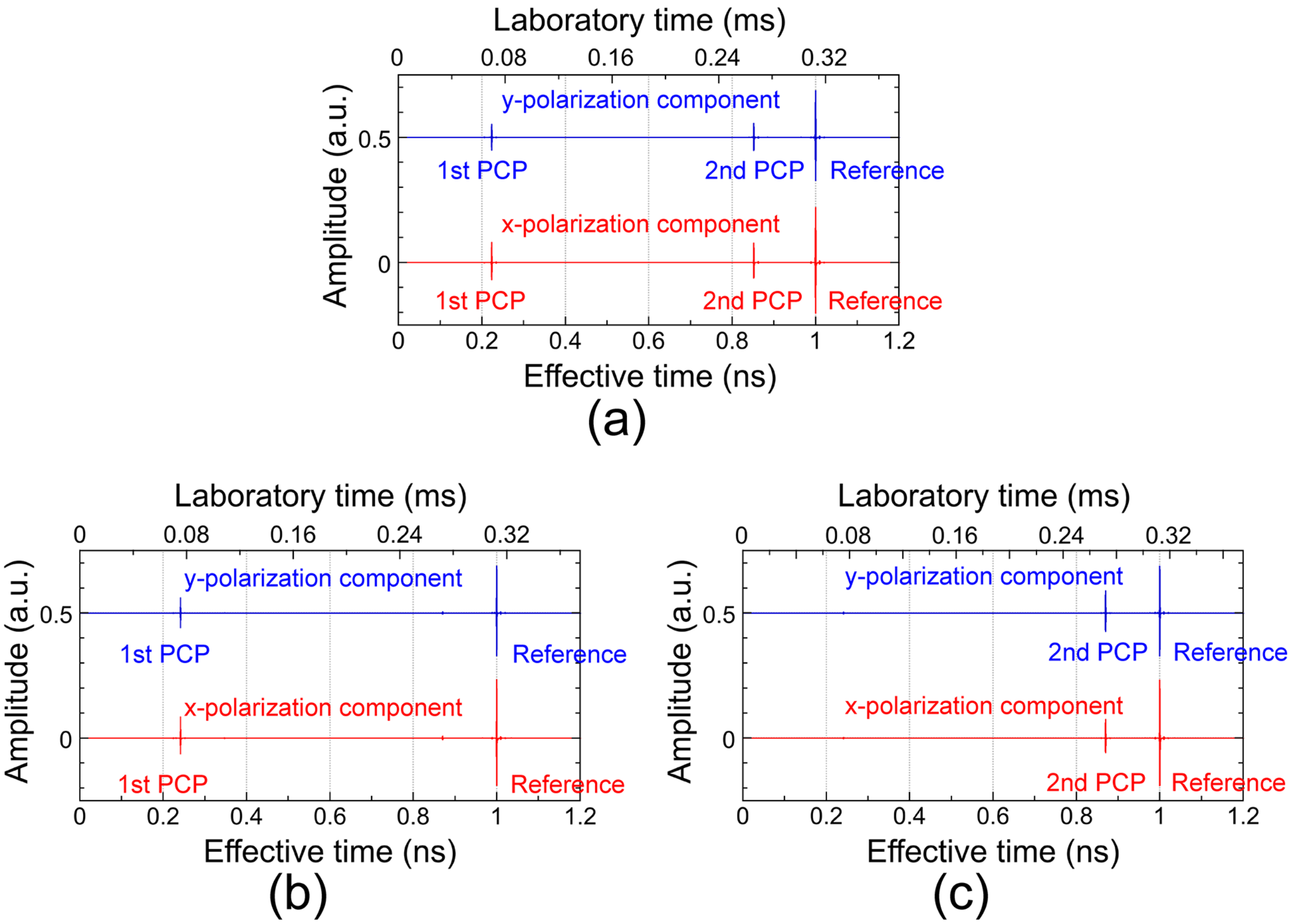

Fig. 4. (a) Interferogram of no samples with respect to x-polarization and y-polarization components. Interferogram with respect to x-polarization and y-polarization components when a transmission axis of the polarizer was orientated at an angle of (b) +45° and (c) +135°.

### *3.2 Measurement of Jones Matrices for Standard Optical Components*

We measured polarization elements with known polarization property to demonstrate the effectiveness of JM-DCSP. We used a zero-order quarter-wave plate (Thorlabs, WPQ05M-1550, wavelength = 1550 nm, retardance accuracy < λ/300) as the first sample with a known birefringence. We averaged 10,000 interferograms to achieve moderate SNR in the following experiments, corresponding to the acquired time of 31 sec; however, it is important to emphasize that the maximum measurement rate can be increased up to $\Delta f_{rep}$ (= 320 Hz). Considering the trade-off between SNR and measurement rate, it is possible to set an appropriate measurement rate for each measurement. Red trace of Fig. 5(a) shows a series of optical spectra of each Jones matrix component in the quarter-wave plate when its fast axis was set to be parallel to the y-polarization or +90°: (a-1) real part and (a-2) imaginary part of $J_{00}$ ($J_{00r}$ and $J_{00i}$), (a-3) real part and (a-4) imaginary part of $J_{01}$ ($J_{01r}$ and $J_{01i}$), (a-5) real part and (a-6) imaginary part of $J_{10}$ ($J_{10r}$ and $J_{10i}$), and (a-7) real part and (a-8) imaginary part of $J_{11}$ ($J_{11r}$ and $J_{11i}$), respectively. Only $J_{00r}$ and $J_{11i}$ are non-zero values whereas the others were zero. This result well reflects the Jones matrix $M_b$ of the birefringent materials provided in the Methods section. For comparison, green trace of Fig. 5(a) shows the corresponding theoretical spectra of each Jones matrix component, which were calculated based on the assumption of a perfect optical element. The comparison between them indicated that the experimental result was in good agreement with the theoretical value for each Jones matrix component.

Each component of Jones matrix in the quarter-wave plate depends on the angle of the fast axis in addition to the birefringence (see the Appendix section). To confirm such the angle dependence of Jones matrix component, the fast axis of the quarter-wave plate was rotated by +40° in a counterclockwise direction from the orientation along the y-axis, corresponding to an angle of +130°. Red trace of Fig. 5(b) shows a series of optical spectra in each Jones matrix component: (b-1) $J_{00r}$, (b-2) $J_{00i}$, (b-3) $J_{01r}$, (b-4) $J_{01i}$, (b-5) $J_{10r}$, (b-6) $J_{10i}$, (b-7) $J_{11r}$, and (b-8) $J_{11i}$. Those spectra were different from that in Fig. 5(a) due to the angle dependence. For comparison, green trace of Fig. 5(b) shows the corresponding theoretical spectra of each Jones matrix component. The experimental result was in moderate agreement with the theoretical value for each Jones matrix component, again.

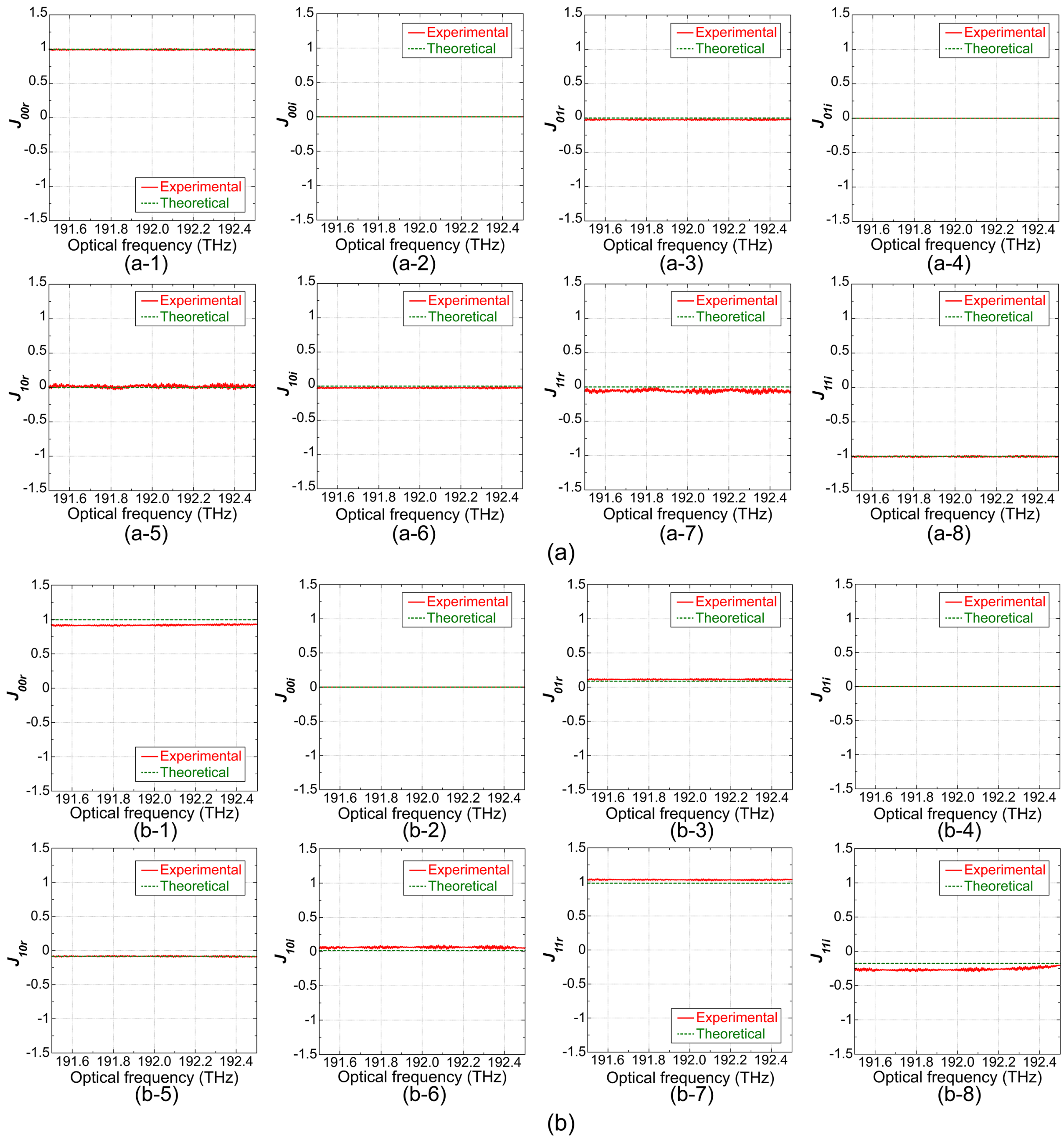

Fig. 5. Optical spectra of eight Jones matrix components in a zero-order quarter-wave plate when its fast axis is parallel to the y-polarization: (a-1) $J_{00r}$, (a-2) $J_{00i}$, (a-3) $J_{01r}$, (a-4) $J_{01i}$, (a-5) $J_{10r}$, (a-6) $J_{10i}$, (a-7) $J_{11r}$, and (a-8) $J_{11i}$. Optical spectra of eight Jones matrix components in a zero-order quarter-wave plate when its fast axis is orientated at an angle of +130°: (b-1) $J_{00r}$, (b-2) $J_{00i}$, (b-3) $J_{01r}$, (b-4) $J_{01i}$, (b-5) $J_{10r}$, (b-6) $J_{10i}$, (b-7) $J_{11r}$, and (b-8) $J_{11i}$.

We also measured a multi-order quarter-wave plate (Thorlabs, WPMQ05M-1550, wavelength = 1550 nm, retardance accuracy < λ/300, the order of the multi-order is not disclosed) when its fast axis was set to be parallel to the y-polarization. Red trace of Fig. 6 shows a series of optical spectra in each Jones matrix component: (a) $J_{00r}$, (b) $J_{00i}$, (c) $J_{01r}$, (d) $J_{01i}$, (e) $J_{10r}$, (f) $J_{10i}$, (g) $J_{11r}$, and (h) $J_{11i}$. For comparison, green trace of Fig. 6 shows the corresponding theoretical spectra of each Jones matrix component, which was interpolated from the product datasheet. Although the spectral bandwidth was limited by BPF, we confirmed the wavelength dependence in $J_{11r}$ and $J_{11i}$ enhanced by large birefringence in the multi-order quarter-wave plate compared with the zero-order quarter-wave plate in Figs. 5(a) and 6(b).

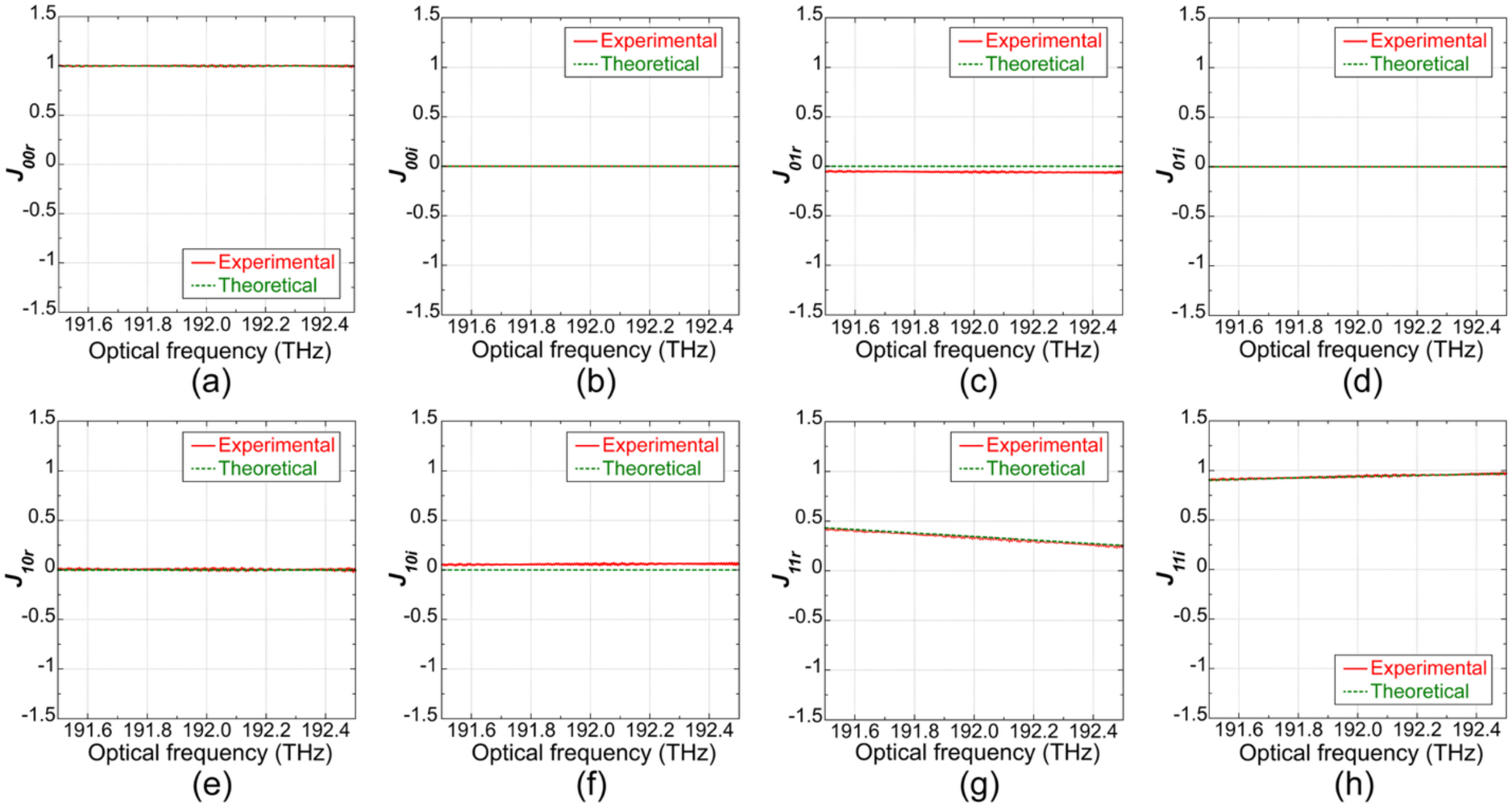

Fig. 6. Optical spectra of eight Jones matrix components in multi-order quarter-wave plate when its fast axis was parallel to the y-polarization: (a) $J_{00r}$, (b) $J_{00i}$, (c) $J_{01r}$, (d) $J_{01i}$, (e) $J_{10r}$, (f) $J_{10i}$, (g) $J_{11r}$, and (h) $J_{11i}$.

We next measured Jones matrix of a Faraday rotator (Thorlabs, I1550R5, rotation angle = 45±3°, minimum transmission = 98%, wavelength = 1500-1600 nm) as a sample of optically active material. When the first PCP with +45° linear polarization and the 2nd PCP with +135° linear polarization were used for the incident light of JM-DCSP, the first and second PCPs after passing through the Faraday rotator have the y-polarization and x-polarization. This situation is corresponding to the dead zone of JM-DCSP. To avoid it, we changed the polarization angle of the first PCP and the second PCP to be +32° and +122°, respectively. Figure 7 compares the experimental data (red trace) with the theoretical value (green trace) regarding a series of optical spectra in each Jones matrix component: (a) $J_{00r}$, (b) $J_{00i}$, (c) $J_{01r}$, (d) $J_{01i}$, (e) $J_{10r}$, (f) $J_{10i}$, (g) $J_{11r}$, and (h) $J_{11i}$. Those theoretical curves were calculated based on the assumption of a perfect optical element. A unique Jones matrix that is different from the quarter-wave plates (see Figs. 5 and 6) can be confirmed. The experimental data and the theoretical value were in good agreement with each other, again. Furthermore, when the Faraday rotator was rotated by +60°, the optical spectra of each Jones matrix component did not change because of no angle dependence of optical rotation (not shown).

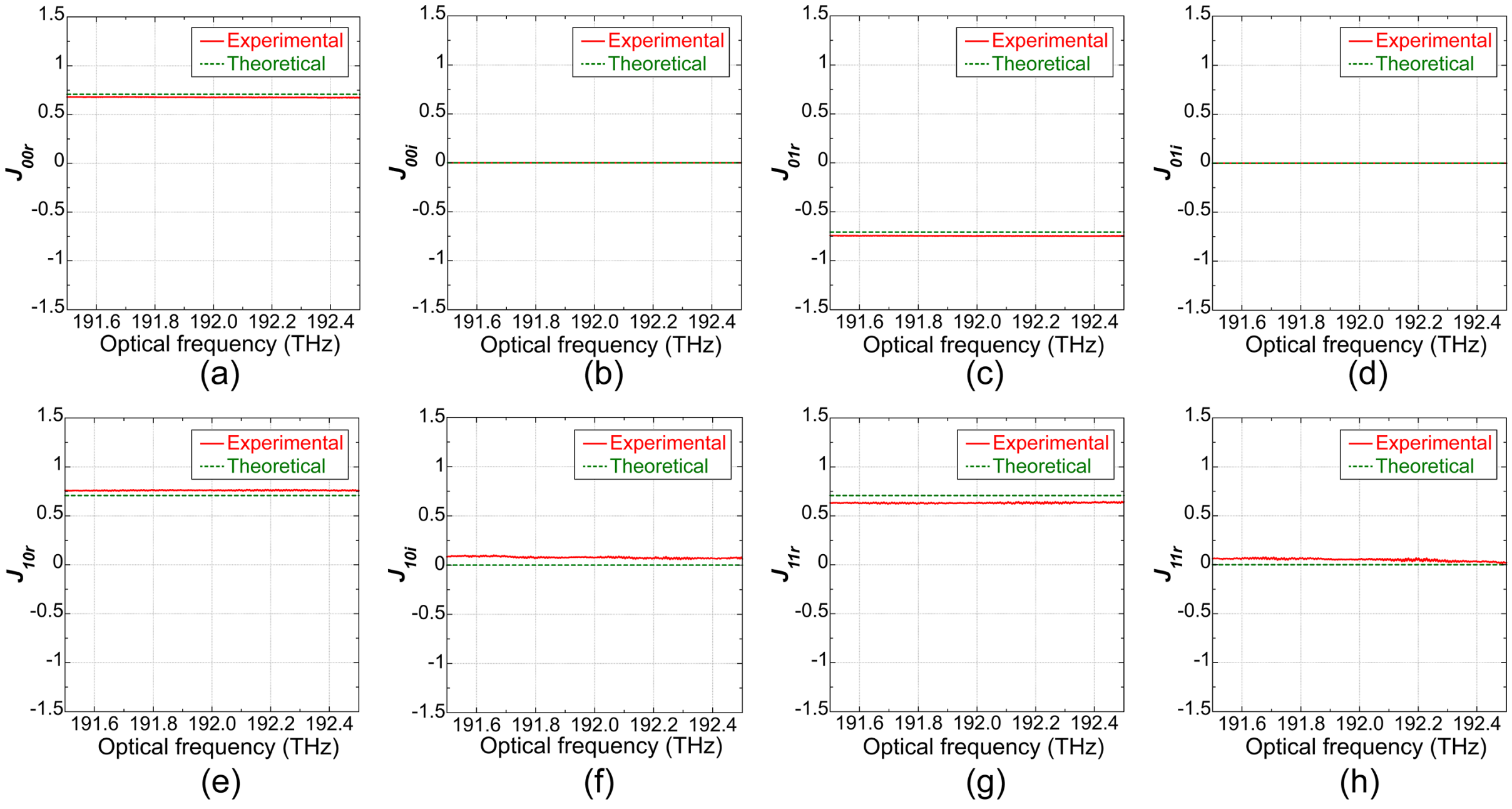

Fig. 7. Optical spectra of eight Jones matrix components in a Faraday rotator: (a) $J_{00r}$, (b) $J_{00i}$, (c) $J_{01r}$, (d) $J_{01i}$, (e) $J_{10r}$, (f) $J_{10i}$, (g) $J_{11r}$, and (h) $J_{11i}$.

Finally, we measured the combined Jones matrix for the series configuration of a Faraday rotator and a zero-order quarter-wave plate. In this case, the combined Jones matrix can be expressed as the product of the Jones matrix of the Faraday rotator and the that of the quarter-wave plate. We here set the fast axis of the quarter-wave plate to be an angle of +130°. Red trace of Fig. 8 shows a series of measured optical spectra in each component of the combined Jones matrix: (a) $J_{00r}$, (b) $J_{00i}$, (c) $J_{01r}$, (d) $J_{01i}$, (e) $J_{10r}$, (f) $J_{10i}$, (g) $J_{11r}$, and (h) $J_{11i}$. These spectra well reflect the product of Jones matrices spectra in Figs. 5(b) and 7. Compared with their theoretical values as shown by green trace in Fig. 8, we confirmed the validity of the proposed method to the combined Jones matrix.

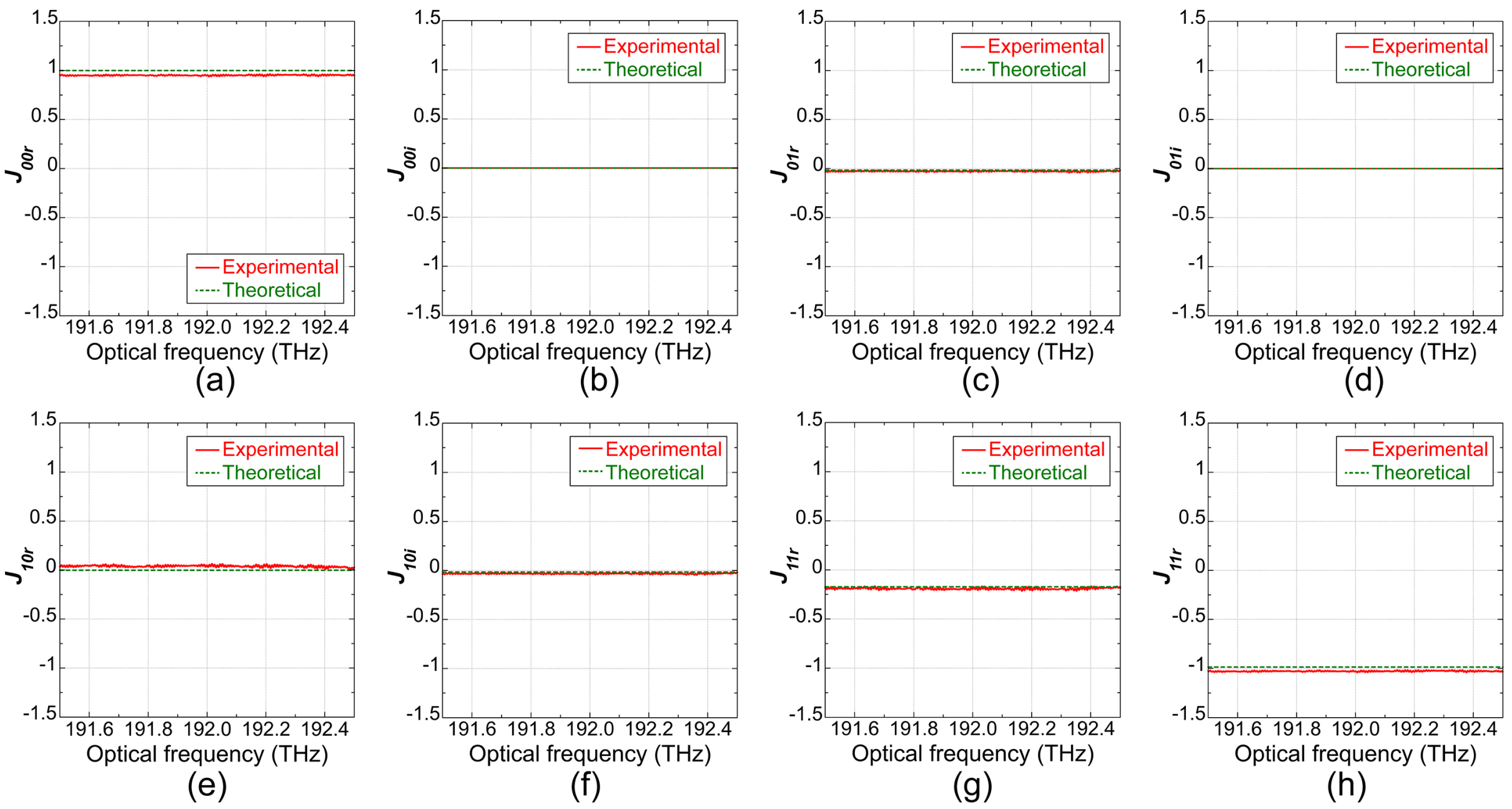

Fig. 8. Optical spectra of the combined Jones matrix in a Faraday rotator and a zero-order quarter-wave plate orientated at 130°: (a) $J_{00r}$, (b) $J_{00i}$, (c) $J_{01r}$, (d) $J_{01i}$, (e) $J_{10r}$, (f) $J_{10i}$, (g) $J_{11r}$, and (h) $J_{11i}$.

## 4 Discussion

The experimental results obtained through JM-DCSP show good agreement with theoretical expectations, but some discrepancies were observed. These discrepancies are attributed to systematic errors that may originate from small polarization crosstalk in the PBSs, residual birefringence in the BSs, and slight misalignment of the optical axes in individual optical components, which can lead to imperfect separation of the x- and y-polarization components. To minimize such systematic errors, calibration of the entire JM-DCSP system using standard optical samples with well-defined polarization properties will be necessary. In addition to these systematic errors, small variations due to random errors were also observed in the experimental results. Random errors arise primarily from environmental disturbances such as air turbulence, mechanical vibrations, and temperature fluctuations. These factors cause random variations in the refractive index of air and minute changes in optical pathlengths, which induce phase drifts in the interferograms and, consequently, in the reconstructed Jones matrix elements. This effect is particularly pronounced because each PCP and the reference pulse are processed independently in separate optical paths. Although efforts were made to mitigate these disturbances by enclosing the optical systems in a plastic box, these combined effects inevitably introduced phase noise and amplitude imbalance. Future improvements, such as the implementation of polarization-maintaining single-mode fiber optical paths and active path-length stabilization,[25] are expected to mitigate these errors and enhance phase stability. Another source of random error is the noise characteristics of OFC sources. The intensity noise of the OFCs can cause amplitude fluctuations between successive interferograms, slightly degrading SNR of the reconstructed Jones-matrix spectra. The phase noise of the OFCs leads to fluctuations in the optical phase of each comb mode, resulting in random phase drifts that introduce small errors in the retrieved phase components of the Jones-matrix elements. However, since both OFCs used in JM-DCSP are tightly phase-locked to each other and the interferograms are coherently averaged over many repetitions, the impact of these noises is largely suppressed, and only minor residual deviations are observed.

To enhance the functionality of DCSP, we introduced a PCPS scheme and validated its effectiveness for standard optical components. By controlling both polarization states and time delays, PCPS provides enhanced flexibility in polarization analysis, allowing for more precise measurements over a broad wavelength range. However, as mentioned in the previous experiment, we identified the dead zone when using two orthogonal-polarized PCPS. The dead zone occurs when the polarization of the transmitted light through the sample matches the x- or y-polarization component. This results in a lack of information in certain regions of the measurement. To eliminate this dead zone, two complementary approaches can be considered. First, combining the present linear basis with alternative orthonormal polarization bases, such as right/left circular[19)] or elliptical[20)] polarization pairs, could effectively remove or reduce the dead zone. Second, further multiplexing of the linear polarization states within the PCPS could enable more comprehensive coverage of the Jones matrix elements. Although these approaches might introduce additional complexity into the optical system, the benefits in terms of improved measurement fidelity and expanded measurement ranges would outweigh the added complexity. Additionally, the increased flexibility offered by a more multiplexed PCPS could contribute to further advancements in high-precision SP by enabling more comprehensive polarization analysis, ultimately reducing errors and increasing reliability in precise and/or dynamic time-resolved measurements.

Although a direct comparison with existing SP measurement techniques was not conducted in this work, several key advantages of the JM-DCSP method are inferred based on previous research.

**Speed:** The measurement rate of JM-DCSP is dependent on the repetition rate difference ($\Delta f_{rep}$) between the two OFCs. Practically, this allows for measurement rates on the order of sub-kHz to kHz, which is significantly higher than traditional methods that rely on mechanical polarization modulation. The high-speed capability of DCSP enables real-time or dynamic polarization analysis, which is a significant advantage for applications requiring rapid data acquisition.[9,10] The advantage of high speed is maintained even when transitioning from DCSP with a single polarization pulse to JM-DCSP with PCPS.

**Spectral resolution:** While the current experiment does not achieve the highest possible spectral resolution, significant potential exists for improving this through optimization of the interferogram acquisition. By adjusting the optical pathlengths, the interferograms from PCPS and the reference pulse could be more evenly spaced within one cycle, enabling spectral resolution that is determined by multiplying $f_{rep1}$ by the number of pulses (in this case, three). This approach could offer a spectral resolution that is orders of magnitude higher than that of traditional SP, which are typically limited by the dispersive or Fourier-transform spectrometer.

**Polarization analysis performance:** One of the key benefits of JM-DCSP is its ability to perform polarization analysis without the need for polarization modulation. This capability has already been demonstrated in previous studies of DCSP, where it was effective in precise analysis of a thin film.[8,10] This advantage should still be preserved even when transitioning from DCSP with a single polarization pulse to JM-DCSP with PCPS, similar to the high-speed capability.

These differences are summarized in Table 1, which highlights the clear advantages of JM-DCSP over conventional SP methods in terms of speed, spectral resolution, and polarization-modulation-free operation. Collectively, these features demonstrate that JM-DCSP offers superior performance and reliability for polarization analysis across a broad range of applications, from materials characterization to real-time biological sample monitoring.

Table 1. Comparison between conventional spectroscopic polarimetry (SP) and the proposed Jones-matrix dual-comb spectroscopic polarimetry (JM-DCSP).

| Category | Conventional SP (polarization-modulation-based) | Proposed JM-DCSP (polarization-modulation-free) |
|---|---|---|
| **Principle** | Polarization analysis using mechanical or electro-optical polarization modulation combined with dispersive or Fourier-transform spectrometers. | Polarization analysis based on dual-comb interferometry combined with PCPS. |
| **Measurement Speed** | Typically limited to a few Hz–tens of Hz due to mechanical polarization modulation. | Determined by the repetition rate difference ($\Delta f_{rep}$), typically sub-kHz to kHz, enabling real-time or dynamic measurements. |
| **Spectral Resolution** | Limited by the resolution of grating or Fourier-transform spectrometers. | Potentially orders of magnitude higher; determined by $f_{rep1}$ × number of pulses, tunable via optical pathlength optimization. |
| **Mechanical Stability** | Sensitive to mechanical vibration and polarization modulator instability. | No moving parts; inherently stable due to interferometric detection with phase-locked OFCs. |
| **Polarization Modulation** | Required (mechanical or electro-optical). | Not required; polarization is controlled by PCPS without modulation. |
| **Typical Applications** | Thin-film characterization, birefringence or optical activity measurement under static conditions. | High-speed and precise polarization analysis of dynamic samples, e.g., real-time material monitoring or bio-imaging. |
| **Key Advantages** | Well-established technique with simple implementation. | High speed, high resolution, and excellent stability with simultaneous amplitude and phase retrieval. |

## 5 Conclusion

In this article, we introduced Jones matrix dual-comb spectroscopic polarimetry (JM-DCSP), a groundbreaking technique that combines dual-comb spectroscopic polarimetry (DCSP) with polarization control pulse sequences (PCPS) to provide a comprehensive analysis of the polarization response of optical samples across a wide wavelength range. Unlike conventional spectroscopic polarimetry (SP), which relies on mechanical polarization modulation and suffers from issues such as instability and slow data acquisition, JM-DCSP removes the need for mechanical modulation, providing a more stable, faster, and higher-resolution solution. Additionally, the use of diverse PCPS enhances the functionality and versatility of SP analysis. With its polarization-modulation-free approach as well as comprehensive analysis of Jones matrix elements, JM-DCSP offers a significant leap forward in SP, positioning it as a powerful tool for a variety of scientific and industrial applications.

Our experimental results demonstrated the effectiveness of JM-DCSP in accurately capturing the polarization properties of optical materials, such as quarter-wave plates and Faraday rotators, with good agreement between experimental data and theoretical models. The rapid acquisition of polarization spectra combined with high precision and the ability to obtain both the real and imaginary parts' spectra of the Jones matrix, positions JM-DCSP as a versatile tool for advanced optical characterization. Looking ahead, further developments in JM-DCSP could focus on reducing environmental disturbances by integrating fiber-optic-based system with enhanced active stabilization, enhancing the robustness of the system and enable even higher precision measurements. Moreover, further multiplexing PCPS could eliminate the dead zone and provide more comprehensive coverage of the Jones matrix elements, thereby improving measurement fidelity and extending the technique's capability for dynamic, high-precision polarization analysis.

The demonstrated versatility of JM-DCSP further extends its potential across a wide range of applications, from real-time monitoring of dynamic phenomena to precise materials characterization. With its ability to analyze complex optical properties with high speed and accuracy, JM-DCSP is poised to drive future advancements in both basic and applied photonics research. Its broad applicability, including in fields ranging from biomedical imaging[26] to nanotechnology and advanced materials science, underscores its value as a powerful tool for understanding light-matter interactions and enabling groundbreaking research in diverse scientific and industrial areas.

## 6 Appendix: Materials and Methods

### *6.1 Jones calculus*

The polarization state of light is determined by the optical amplitudes and phases of the orthogonal x-polarized component and y-polarized component in electric field. This polarization state is represented using the Jones vector $\boldsymbol{E}$, expressed by

$$\boldsymbol{E} = \begin{bmatrix} E_x \cdot e^{i\delta_x} \\ E_y \cdot e^{i\delta_y} \end{bmatrix}, \tag{9}$$

where $E_x$ and $E_y$ respectively represent the amplitudes of the x-polarized component and y-polarized component whereas $\delta_x$ and $\delta_y$ represent the phases of the x-polarized component and y-polarized component, respectively.

Next, polarization property of a sample is represented using the Jones matrix $\boldsymbol{J}$ by

$$\boldsymbol{J} = \begin{bmatrix} J_{11} & J_{12} \\ J_{21} & J_{22} \end{bmatrix}, \tag{10}$$

where $J_{11}$, $J_{12}$, $J_{21}$, and $J_{22}$ are the complex elements of the Jones matrix. For example, the Jones matrices for birefringent material ($\boldsymbol{J_b}$) and optically active material ($\boldsymbol{J_r}$), respectively, are represented by

$$\boldsymbol{J_b} = \begin{bmatrix} 1 & 0 \\ 0 & e^{ib} \end{bmatrix} \tag{11}$$

$$\boldsymbol{J_r} = \begin{bmatrix} cos\ r & -\sin r \\ \sin r & \cos r \end{bmatrix} \tag{12}$$

where $b$ and $r$ represent the phase differences due to birefringence and the optical rotation angle, respectively. $\boldsymbol{J_b}$ with the polarization axis rotated by an angle $\theta$ from the x-axis is represented as $\boldsymbol{J_b(\theta)}$ by

$$\boldsymbol{J_b}(\boldsymbol{\theta}) = \begin{bmatrix} cos\theta & -sin\theta \\ sin\theta & cos\theta \end{bmatrix} \cdot \boldsymbol{J_b} \tag{13}$$

After passing through a sample given by $\boldsymbol{J}$, the polarization state of incident light given by $\boldsymbol{E}$ is transformed to a polarization state of output light given by $\boldsymbol{E'}$. This transformation can be expressed using the Jones calculus by

$$\boldsymbol{E'} = \boldsymbol{J} \cdot \boldsymbol{E}. \tag{14}$$

## Code and Data Availability

Data underlying the results presented in this paper are not publicly available at this time but may be obtained from Takeshi Yasui upon reasonable request.

*Acknowledgments*

The authors acknowledge financial support from Forming Japan's Peak Research Universities (J-PEAKS, Grant No. JPJS00420240022) and from Grants-in-Aid for Scientific Research (Grant Nos. 20J23577, 22H00303, and 24K21237), both funded by the Japan Society for the Promotion of Science (JSPS). Additional support was provided by the Cabinet Office, Government of Japan (Promotion of Regional Industries and Universities), Tokushima Prefecture (Creation and Application of Next-Generation Photonics), and the Research Clusters Program of Tokushima University (2201001). In accordance with the journal guidelines, the use of the large language model ChatGPT for refining language and grammar is acknowledged.

*References*

1. H. Fujiwara, *Spectroscopic ellipsometry: principles and applications*, John Wiley & Sons, Hoboken (2007).

2. C. Yim et al., “Investigation of the optical properties of $MoS_2$ thin films using spectroscopic ellipsometry,” *Appl. Phys. Lett.* **104**(10), 103114 (2014) [doi:10.1063/1.4868108].
3. S. G. Lim et al., “Dielectric functions and optical bandgaps of high-K dielectrics for metal-oxide-semiconductor field-effect transistors by far ultraviolet spectroscopic ellipsometry,” *J. Appl. Phys.* **91**(7), 4500-4505 (2002) [doi: 10.1063/1.1456246].
4. K. Spaeth et al., “Studies on the biotin-avidin multilayer adsorption by spectroscopic ellipsometry,” *J. Colloid Interface Sci.* **196**(2), 128-135 (1997) [doi:10.1006/jcis.1997.5200].
5. D. E. Aspnes, “Fourier transform detection system for rotating-analyzer ellipsometers,” *Opt. Commun.* **8**(3), 222-225 (1973) [doi:10.1016/0030-4018(73)90132-6].
6. P. S. Hauge et al., “A rotating-compensator Fourier ellipsometer,” *Opt. Commun.* **14**(4), 431-437 (1975) [doi:10.1016/0030-4018(75)90012-7].
7. C. Y. Han et al., “Photoelastic modulated imaging ellipsometry by stroboscopic illumination technique,” *Rev. Sci. Instrum.* **77**(2), 023107 (2006) [doi:10.1063/1.2173027].
8. T. Minamikawa et al., “Dual-comb spectroscopic ellipsometry,” *Nat. Commun.* **8**(1), 610 (2017) [doi:10.1038/s41467-017-00709-y].
9. H. Koresawa et al., “Dynamic characterization of polarization property in liquid-crystal-on-silicon spatial light modulator using dual-comb spectroscopic polarimetry,” *Opt. Express* **28**(16), 23584-23593 (2020) [doi: 10.1364/OE.399200].
10. R. Zhang et al., “Dynamic ellipsometry measurement based on a simplified phase-stable dual-comb system,” *Opt. Express* **30**(5), 7806-7820 (2022) [doi:10.1364/OE.453406].
11. K. Hinrichs et al., “Mid-infrared dual-comb polarimetry of anisotropic samples,” *Nat. Sci.* **3**(2), e20220056 (2023) [doi:10.1002/ntls.20220056].
12. Th. Udem et al., “Accurate measurement of large optical frequency differences with a mode-locked laser,” *Opt. Lett.* **24**(13), 881-883 (1999) [doi: 10.1364/OL.24.000881].

13. M. Niering et al., “Measurement of the hydrogen 1S-2S transition frequency by phase coherent comparison with a microwave cesium fountain clock,” *Phys. Rev. Lett.* **84**(24), 5496–5499 (2000) [doi:10.1103/PhysRevLett.84.5496].

14. Th. Udem et al., “Optical frequency metrology,” *Nature* **416**(6877), 233-237 (2002) [doi:10.1038/416233a].

15. S. Schiller, “Spectrometry with frequency combs,” *Opt. Lett.* **27**(9), 766–768 (2002) [doi:10.1364/OL.27.000766].

16. F. Keilman et al., “Time-domain mid-infrared frequency-comb spectrometer,” *Opt. Lett.* **29**(13), 1542–1544 (2004) [doi:10.1364/OL.29.001542].

17. T. Yasui et al., “Terahertz frequency comb by multifrequency-heterodyning photoconductive detection for high-accuracy, high-resolution terahertz spectroscopy,” *Appl. Phys. Lett.* **88**(24), 241104 (2006) [doi:10.1063/1.2209718].

18. I. Coddington et al., “Dual-comb spectroscopy,” *Optica* **3**(4), 414-426 (2016) [doi:10.1364/OPTICA.3.000414].

19. Z. Liu et al., “Dual Jones matrices empowered six phase channels modulation with single-layer monoatomic metasurfaces,” *Laser Photonics Rev.*, **19**(7), 2401526 (2025) [doi:10.1002/lpor.202401526].

20. X. Mu et al., “Chirality-free full decoupling of jones matrix phase-channels with a planar minimalist metasurface,” Nano Lett., **25**(4), 1322-1328 (2025) [doi:10.1021/acs.nanolett.4c04577].

21. A. Nishiyama et al., “Doppler-free dual-comb spectroscopy of Rb using optical-optical double resonance technique,” *Opt. Express* **24**(22), 25894–25904 (2016) [doi:10.1364/OE.24.025894].

22. A. Asahara et al., “Dual-comb spectroscopy for rapid characterization of complex optical properties of solids,” *Opt. Lett.* **41**(21), 4971–4974 (2016) [doi:10.1364/OL.41.004971].

23. E. Baumann et al., “Spectroscopy of the methane $\nu_3$ band with an accurate midinfrared coherent dual-comb spectrometer,” *Phys. Rev. A* **84**(6), 062513 (2011) [doi:10.1103/PhysRevA.84.062513].

24. J. Roy et al., "Continuous real-time correction and averaging for frequency comb interferometry," *Opt. Express* **20**(20), 21932–21939 (2012) [doi:10.1364/OE.20.021932].

25. H. Koresawa et al., "Combination of dual-comb spectroscopy with Jones-matrix polarimetry," in Tech. Digest 15th Pacific Rim Conference on Lasers and Electro-Optics 2022, P-CTh6-09 (2022).

26. X. Dai et al., "Quantitative Jones matrix imaging using vectorial Fourier ptychography," *Biomed. Opt. Express* **13**(3), 1457-1470 (2022) [doi:10.1364/BOE.448804].

**Hidenori Koresawa** was affiliated with Tokushima University when this research was conducted and is currently a postdoctoral scientist at The University of Osaka. He received his BS, MS, and PhD degrees in engineering from Tokushima University in 2018, 2020, and 2023, respectively. His research interests include optical frequency combs.

**Eiji Hase** is an associate professor at Tokushima University. He received his BS, MS, and PhD degrees in engineering from Tokushima University in 2012, 2014, and 2017, respectively. His research interests include biomedical imaging and biomechanical analysis using femtosecond lasers and synchrotron X-rays. He is a member of SPIE.

**Yu Tokizane** is an associate professor at Tokushima University. He received his BS degree in engineering from Hokkaido University in 2005, MS degree in engineering from Tokyo Institute of Technology in 2007, and PhD degree in engineering from Hokkaido University in 2010, respectively. His research interests include THz photonics and optical vortex.

**Akifumi Asahara** is an associate professor at The University of Electro-Communications. He received his BS, MS, and PhD degrees in physics from The University of Tokyo in 2009, 2011, and 2015, respectively. His research interests include optical frequency combs and spectroscopy.

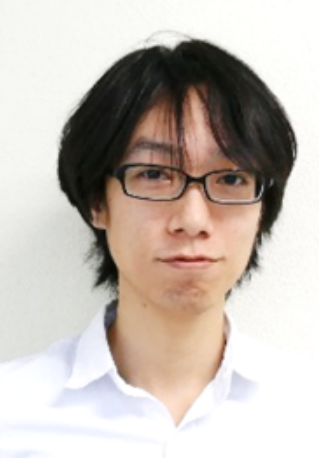

**Takeo Minamikawa** is a professor at The University of Osaka. He received his BS, MS, and PhD degrees in engineering from Osaka University in 2006, 2008, and 2010, respectively. His research interests include Raman microspectroscopy, plasmonics, optical frequency combs and their biomedical applications.

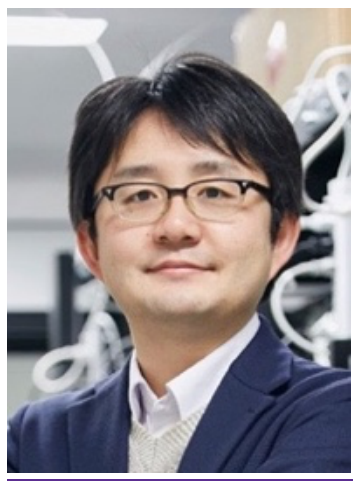

**Kaoru Minoshima** is a professor at The University of Electro-Communications. She received her BS, MS, and PhD degrees in physics from The University of Tokyo in 1987, 1989, and 1993, respectively. Her research interests include optical frequency combs, optical metrology, and spectroscopy. She is Fellows of Optica, SPIE, JSAP, and LSJ.

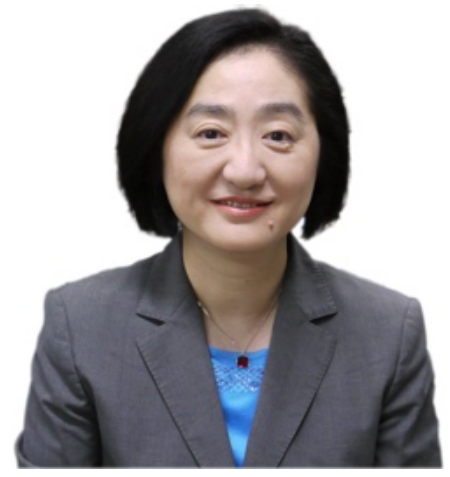

**Takeshi Yasui** is a professor of Tokushima University. He received his first PhD degree in engineering from the Tokushima University in 1997 and his second PhD degree in medical science from Nara Medical University in 2013. His research interests include optical frequency comb, terahertz photonics, and biomedical optics. He is the author of more than 150 journal papers and has written four book chapters. He is a member of SPIE and a senior member of Optica.

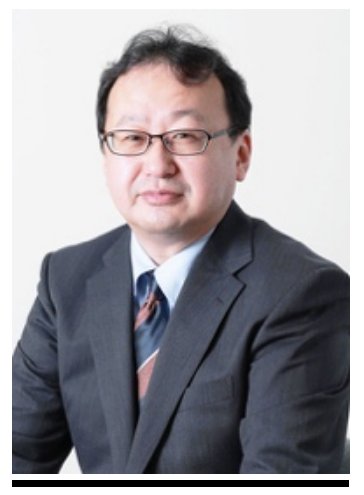

Biographies and photographs for the other authors are not available.

**Caption List**

**Fig. 1** Experimental setup of JM-DCSP. Rb-FS, rubidium frequency standard; λ/4, quarter-wave plate; λ/2, a half-wave plate; PBSs, polarization beam splitters; BSs, beam splitters; P, polarizer; BPF, optical bandpass filter; PDs, photodiodes.

**Fig. 2** Interferogram of no samples with respect to x-polarization and y-polarization components in one repetition period of the interferogram (= 10 ns). Three insets show the temporally magnified interferograms of the 1st PCP, 2nd PCP, and the reference pulse, respectively.

**Fig. 3** Spectral characteristics of the first PCP, the second PCP, and the reference pulse. Amplitude spectra of the x- and y-polarization components of (a) the first PCP, (b) the second PCP, and (c) the reference pulse. (d) Comparison of phase spectra of the x-polarization component among the first PCP, the second PCP, and the reference pulse. Phase difference spectra of the x- and y-polarization components of (e) the first PCP and (f) the second PCP to the reference pulse.

**Fig. 4** (a) Interferogram of no samples with respect to x-polarization and y-polarization components. Interferogram with respect to x-polarization and y-polarization components when a transmission axis of the polarizer was orientated at an angle of (b) +45° and (c) +135°.

**Fig. 5** Optical spectra of eight Jones matrix components in a zero-order quarter-wave plate when its fast axis is parallel to the y-polarization: (a-1) $J_{00r}$, (a-2) $J_{00i}$, (a-3) $J_{01r}$, (a-4) $J_{01i}$, (a-5) $J_{10r}$, (a-6) $J_{10i}$, (a-7) $J_{11r}$, and (a-8) $J_{11i}$. Optical spectra of eight Jones matrix components in a zero-order quarter-wave plate when its fast axis is orientated at an angle of +130°: (b-1) $J_{00r}$, (b-2) $J_{00i}$, (b-3) $J_{01r}$, (b-4) $J_{01i}$, (b-5) $J_{10r}$, (b-6) $J_{10i}$, (b-7) $J_{11r}$, and (b-8) $J_{11i}$.

**Fig. 6** Optical spectra of eight Jones matrix components in multi-order quarter-wave plate when its fast axis was parallel to the y-polarization: (a) $J_{00r}$, (b) $J_{00i}$, (c) $J_{01r}$, (d) $J_{01i}$, (e) $J_{10r}$, (f) $J_{10i}$, (g) $J_{11r}$, and (h) $J_{11i}$.

**Fig. 7** Optical spectra of eight Jones matrix components in a Faraday rotator: (a) $J_{00r}$, (b) $J_{00i}$, (c) $J_{01r}$, (d) $J_{01i}$, (e) $J_{10r}$, (f) $J_{10i}$, (g) $J_{11r}$, and (h) $J_{11i}$.

**Fig. 8** Optical spectra of the combined Jones matrix in a Faraday rotator and a zero-order quarter-wave plate orientated at 130°: (a) $J_{00r}$, (b) $J_{00i}$, (c) $J_{01r}$, (d) $J_{01i}$, (e) $J_{10r}$, (f) $J_{10i}$, (g) $J_{11r}$, and (h) $J_{11i}$.

**Table 1** Comparison between conventional spectroscopic polarimetry (SP) and the proposed Jones-matrix dual-comb spectroscopic polarimetry (JM-DCSP).